\documentstyle[11pt,moriond,epsfig]{article}

\def\be{\begin{equation}}
\def\ee{\end{equation}}
\def\bea{\begin{eqnarray}}
\def\eea{\end{eqnarray}}

\begin{document}
\title{TWO ELECTRON VIEW ON METAL-INSULATOR TRANSITION\\
 IN TWO DIMENSIONS}

\author{ D. L. SHEPELYANSKY }

\address{Laboratoire de Physique Quantique, UMR 5626 du CNRS,\\
Universit\'e Paul Sabatier, F-31062 Toulouse Cedex 4, France}

\maketitle\abstracts{
The model of two electrons with Coulomb interaction on a two-dimensional
(2D) disordered lattice is considered. It is shown that
the interaction can give a sharp transition to delocalized states
in a way similar to the Anderson transition in 3D.
The localized phase appears when the ratio of the Coulomb energy
to the Fermi energy becomes larger than some critical value 
dependent on the disorder. The relation to the experiments
on metal-insulator transition in 2D is also discussed.}

\section{Introduction}
According to D. Tsui \cite{tsui}, ``the important thing is the 
interplay between
disorder and electron-electron interactions. The FQHE 
(fractional quantum Hall effect)
is, in some sense, the clean limit. But there's another limit,
where both interaction and disorder are important  ... 
there's always some disorder.''
Indeed, the recent experimental discovery of metal-insulation
transition in two dimensions (2D) by Kravchenko {\it et al.} 
\cite{krav94} attracted a great interest to this problem.
This transition is especially surprising since according to
the well established theoretical
result \cite{and79} all states of non-interacting
electrons in 2D disordered potential are exponentially localized.
However, in reality the electron-electron interaction is 
present and the original result \cite{krav94}, as well as the new results
of different groups in experiments with different materials
\cite{krav96,popovic,canada,yael,alex,pudalov,ensslin}, show that
the interaction can induce metallic behavior. Indeed, the majority
of experiments are done in the situation where the parameter
$r_s = 1/{\sqrt{\pi n_s} a^*_{B}} \simeq E_{ee}/E_F \gg 1$.
Here, $E_{ee}$ is the energy of Coulomb interaction,
$E_F$ is the Fermi energy determined by the charge density $n_s$
and $ a^*_{B}$ is the effective Bohr
radius. In some experiments the $r_s$ value was as large as 10 - 30.
In this situation the electrons are located far from each other
and in a first approximation it is natural to consider
the problem of only two electrons with Coulomb interaction.
The first consideration of
two particles with strong attraction was done by Dorokhov
\cite{dorokhov} but it was ignored by the community.
The studies of two interacting particles with short range interaction 
showed that repulsive/attractive interaction
can lead to a strong increase of localization length
or even to delocalize pairs of particles in dimension $d>2$
\cite{ds94,imry,pichard,oppen,moriond}.  According to \cite{ds94,imry} 
in 2D the pairs
of particles remain localized and their localization length
$l_c$ grows smoothly with the increase of disorder strength $U$
or one-particle localization length $l_1$:
$\ln (l_c/l_1) \sim \kappa >$ with $\kappa \sim \Gamma_2 \rho_2$,
where $\Gamma_2 \sim U^2/Vl_1^2$ is the interaction induced transition rate, 
$\rho_2 \sim l_1^4/V$ is
two-particle density of states in the middle of the band
and $V$ is the hopping strength proportional to the energy band size $B$
($B=4V$ for weak disorder). The case of the long range Coulomb interaction
requires separate analysis. Generally, one can expect that the delocalization 
effect will be even stronger in this case since 
the particles are always interacting, in a difference from 
a short range interaction case.

\section{Analytical estimates}
The first estimates for two electrons with Coulomb interaction on a  2D
disordered Anderson lattice were presented in \cite{ds99}.
The lattice is characterized by the nearby hopping $V$
and the diagonal disorder in the interval 
$[-W/2,W/2]$, while the interaction is $U/|r_1-r_2|$.
Then the parameter $r_s=U/(2V\sqrt{\pi n_s})$ where $n_s$ is the filling factor.
If the distance $R$ between the electrons is much larger than 
the one-particle localization length $l_1$ ($\ln l_1 \sim (V/W)^2$)
then the 
\begin{figure}
\hskip 0.5in
\psfig{figure=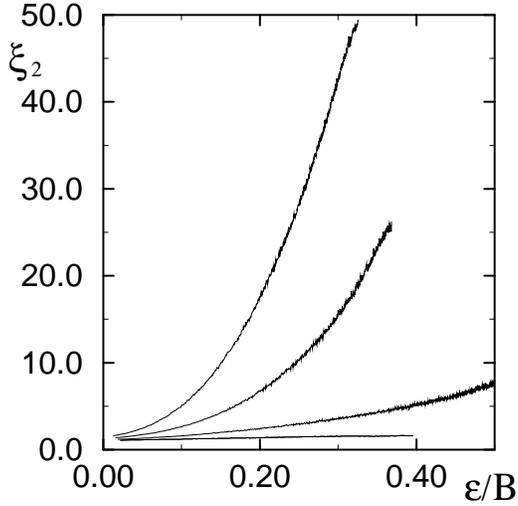,height=3.0in}
\caption{Dependence of the inverse participation ratio $\xi_2$
on rescaled one-particle energy $\epsilon/B$
for $U/V=~2,\;\; L=~16 \; (r_s=6.38, B=4V)$ and $W/V=5, 7, 10, 15 \;$ 
(from up to down).
}
\end{figure}
two-body coupling appears only in the dipole-dipole interaction
term.  This gives the typical matrix element $U_s \sim U/R^3$ \cite{ds99}
and the transition rate $\Gamma_2 \sim U_s^2 \rho_2$, where
still $\rho_2$ is determined by the estimate given above since
the electrons can have a jump only on a distance $l_1$
from the initial position (otherwise wave function overlap
drops exponentially). As the result the two electron levels
become mixed by interaction when $\kappa_e = \chi_e^2 \sim \Gamma_2 \rho_2
\sim (r_L^{4/3}/r_s)^2 > 1$, where $r_L$ is the value of $r_s$
at the density $n_s=1/l_1^2$ (one electron in a box of $l_1$ size).
For $\chi_e > 1$ the Coulomb interaction leads to a delocalization 
of two electrons in a way similar to 3D Anderson transition\cite{ds99}.
Indeed, in this case the hopping goes effectively in 3D:
the center of mass moves in 2D and in addition the electrons
slowly rotate around it that gives 3 dimensions.
The rotation goes on a ring of width $l_1$ and of radius 
$R \sim l_1^{4/3} \gg l_1$ (for $U \sim V$);
the size of the ring is fixed by the energy conservation
$\epsilon \sim U/R$. Due to that 
the length $l_c$ changes sharply from $l_c \sim l_1$
to $l_c \sim l_1 \exp(\pi l_1^{1/3} \kappa_e)$ when $\kappa_e$ crosses
the critical value $\kappa_e \sim 1$ \cite{ds99}. 
It is interesting to note that,
as in the experiments (see Refs. 2,4-10),
the localized phase corresponds to the large
values of $r_s$: physically the two-body interaction becomes weaker
at low density. The diffusion rate in the metallic phase can be estimated
as $D_e \sim l_1^2 \Gamma_2 \sim V \kappa_e/l_1^2$. Near the critical 
point $\kappa_e \sim 1$ the diffusion rate (conductivity)
drops with the decrease of disorder (increase of $l_1$).
These estimates are done for the excited states in the middle of the band.

\section{Numerical results}
The above problem of two electrons in the 2D Anderson model
in the triplet state is studied numerically. The maximal lattice size
is $L=24$. The numerical diagonalization is done in the following way:
the Hamiltonian is rewritten in the basis of noninteracting
eigenstates, from which only first $M$ low energy one-particle states 
(orbitals) are selected and after that the Hamiltonian is diagonalized 
exactly. The special check is done to ensure that the low energy states
are not effected by the above cutoff (e.g. by changing $M$ in few times).
Usually ND=4000 disorder realizations are used to average the fluctuations.
The fact that the effect of interaction strongly depends on $l_1$
(or $W$) is demonstrated in Fig. 1. Indeed, here the number
of noninteracting eigenstates $\xi_2$ contributing in 
an eigenfunction at fixed interaction $U/V=2$ is increased
in about 50 times only by the change of the disorder $W$.
This confirms the analytical result according to which
the effect of interaction becomes stronger for larger $l_1$
since $\xi_2 \sim \Gamma_2 \rho_2(\epsilon)$. 
\begin{figure}
\hskip 1.5in
\psfig{figure=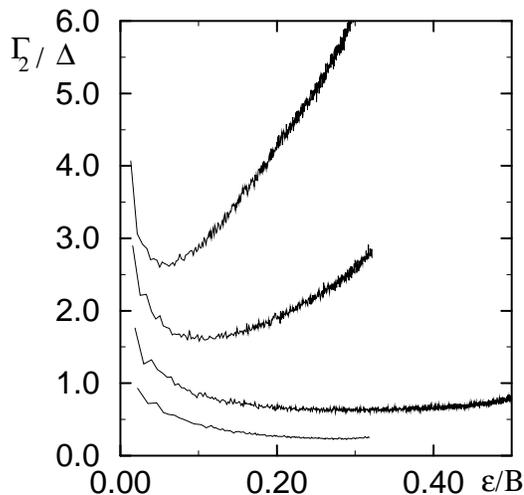,height=3.0in}
\caption{Dependence of the rescaled transition rate $\Gamma_2$
(defined via the relation $\Gamma_2 = (\xi_2-1)/ \rho_2 (\epsilon)$
with $\rho_2 (\epsilon)$ being the two-electron density of states)
on $\epsilon/B$ for the parameters of Fig.1 and the same order
of curves, $\Delta=B/L^2$.
}
\end{figure}
The last relation
allows to determine numerically the dependence of the transition rate
$\Gamma_2$ on the excitation energy
$E=2 \epsilon$, counted from the ground state. This dependence is 
presented in Fig. 2 and for a moderate disorder shows a very
flat dependence on $\epsilon$ and even a certain increase
of $\Gamma_2$ very close to the ground state. If to assume that
the matrix element $U_s$ is independent of $\epsilon$ 
then $\Gamma_2$ should drop linearly with $\epsilon$ since
$\rho_2(\epsilon) \approx l_1^4 \epsilon/V^2$ \cite{imry}.
However, for a short range interaction is has been shown that
for localized states $\Gamma_2$ can be independent of $\epsilon$
due to enhanced return probability near the Fermi level
\cite{jack}. For the case of long range interaction similar effects can be
responsible for the flat variation of $\Gamma_2$ with $\epsilon$ 
in Fig. 2. More detailed studies are required to understand the properties
of $\Gamma_2$ near the ground state. 

Another part of numerical studies is devoted to the investigation of the
level spacing statistics in the above model. Indeed, it is known
that the localized phase is characterized by the Poisson distribution
$P_P(s)$,
the metallic phase has the Wigner-Dyson statistics $P_{WD}$ while
the critical transition point has an intermediate statistics
independent of the system size \cite{shklov}. It is convenient 
to study the transition between two limits with the help
of the parameter $\eta=\int_0^{s_0}
(P(s)-P_{WD}(s)) ds / \int_0^{s_0} (P_{P}(s)-P_{WD}(s)) ds$,
where  $s_0=0.4729...$ is the intersection point of $P_P(s)$ and $P_{WD}(s)$.
In this way $\eta=1$
corresponds to $P_P(s)$, and $\eta$=0 to $P_{WD}(s)$. The dependence of $\eta$
on $\epsilon$ is determined in the following way.
For each disorder realization the spacing between nearby energy
levels  $E_i$ is determined and then is averaged over 
ND disorder realizations for each $i$ giving the $P(s)$  statistics and $\eta$
as a function of averaged excitation energy $\epsilon =E/2$.
At higher energies the values of $\eta$ are in addition averaged in a fixed
energy interval. In this way the total statistics obtained
for $P(s)$ and $\eta$ varies from $NS=12000$ for low energy states up to
$NS=10^6$ at high energy with high density of levels. Un example of $\eta$ 
variation with energy for different system size $L$ is shown in Fig. 3
(see also Fig. 1 in \cite{ds99} for a stronger disorder). For large $L$
the statistics becomes close to $P_P(s)$ at low energy
and to $P_{WD}(s)$ at $\epsilon$ larger than a critical energy $\epsilon_c$
dependent on the disorder and independent of $L$. 
The transition in the spectral statistics
can be qualitatively understand on the basis of the estimates
given in the previous section. Indeed, since the interaction
energy is $U/R \sim \epsilon$ the high energy states allow
to have particles closer to each other ($R \sim U/\epsilon$)
that increases their interaction and finally leads to 
delocalization for $R > l_1^{4/3} (U \sim V)$. In agreement with
this picture the critical energy $\epsilon_c$ decreases
with the increase of $l_1$ (decrease of $W/V$) as it can be seen
from Fig. 1 in \cite{ds99} and Fig. 3. The fact that an interaction 
increases the localization length for two particles in 2D has been also
seen in the other numerical simulations \cite{ortuno,cuevas}.
However, the claim made there that the short
range interaction gives a transition from localized to delocalized
states is in a sharp contradiction with the theoretical
arguments \cite{ds94,imry,moriond} and probably should be attributed
to small sizes used in \cite{ortuno,cuevas}. The numerical
data for the Coulomb case presented in \cite{cuevas} are
somewhat similar to the data presented here and in \cite{ds99},
even if any theoretical arguments in the favor of transition were
presented in \cite{cuevas}.

The variation of $P(s)$ with the interaction strength
is shown in Fig. 4. At small $U$ the statistics 
approaches to the Poisson distribution while with the increase
of $U$ it tends to the Wigner-Dyson case. In the vicinity
of the  critical point $\epsilon_c$ the
statistics is close to the critical statistics in the 3D Anderson model
with periodic boundary conditions \cite{shklov,isa,braun} (see Fig. 4).
This gives one more support for the physical picture developed 
in the previous section according to which two electrons 
in 2D are delocalized in a way similar to the 3D Anderson transition.
\begin{figure}
\hskip 1.5in
\psfig{figure=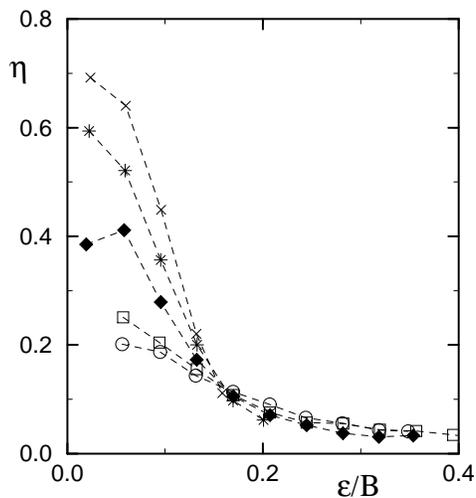,height=3.0in}
\vskip -0.2in
\caption{Dependence of $\eta$ on $\epsilon/B$ for $U/V=2, W/V=5$
and different $L$: 10 (o); 12 ($\Box$); 16 (full diamond); 20 (*); 
24 (x); so that $2.39 \leq r_s \leq 9.57$.
}
\end{figure}

\section{Conclusion}
The analytical and numerical results obtained show that
the Coulomb interaction leads to delocalization of two electron
states in a way similar to the Anderson transition in 3D
for  $r_L < r_s < r_L^{4/3}$.
The model is restricted only by two interacting 
electrons and has delocalization only for
excited states that represents its weak point. 
However, it gives a picture qualitatively
similar to experimentally observed metal-insulator transition in 2D
\cite{krav94} and therefore it can be useful for a future complete theory.

\begin{figure}
\hskip 1.5in
\psfig{figure=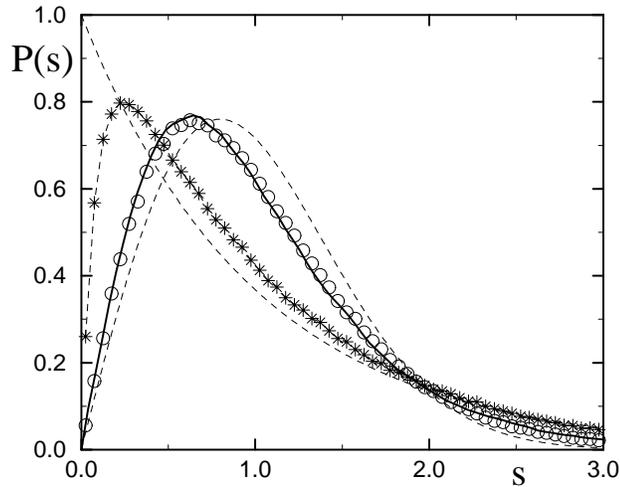,height=3.0in}
\vskip -0.3in
\caption{Level statistics $P(s)$ for two 2D electrons at
$W/V=7,  L=16$ in the energy interval $0.25 \leq \epsilon/B \leq 0.3$:
(o) $U/V=2$, vicinity of the critical point ( see Fig. 1c in Ref. 17);
(*) $U/V=0.2$; total statistics is $NS=5\times 10^5$. 
The full line shows the critical $P(s)$ in 3D Anderson model
$(W/V=16.5, L=14$, taken from Ref. 23); the dashed lines give Poisson
statistics and Wigner surmise.
}
\end{figure}

\end{document}